\documentclass[%
 aip,
rsi,
 reprint,
]{revtex4-1}

\usepackage{graphicx}
\usepackage{dcolumn}
\usepackage{bm}

\usepackage[utf8]{inputenc}
\usepackage[T1]{fontenc}
\usepackage{mathptmx}
\usepackage{etoolbox}

\usepackage[version=4]{mhchem}
\usepackage{color}
\usepackage{amssymb, amsfonts, amsmath}
\usepackage{array}
\usepackage{subcaption}
\usepackage{physics}

\makeatletter
\def\@email#1#2{%
 \endgroup
 \patchcmd{\titleblock@produce}
  {\frontmatter@RRAPformat}
  {\frontmatter@RRAPformat{\produce@RRAP{*#1\href{mailto:#2}{#2}}}\frontmatter@RRAPformat}
  {}{}
}%
\makeatother
\begin{document}

\preprint{AIP/123-QED}

\title{Microwave Hall measurements using a circularly polarized dielectric cavity}

\author{M. Roppongi}
\altaffiliation{The author to whom correspondence may be addressed: roppongi@qpm.k.u-tokyo.ac.jp}
\affiliation{Department of Advanced Materials Science, University of Tokyo, Kashiwa, Chiba 277-8561, Japan}
\email{roppongi@qpm.k.u-tokyo.ac.jp}

\author{T. Arakawa}
\affiliation{National Institute of Advanced Industrial Science and Technology (AIST), Tsukuba, Ibaraki 305-8563, Japan}

\author{Y. Yoshino}
\author{K. Ishihara}
 \affiliation{Department of Advanced Materials Science, University of Tokyo, Kashiwa, Chiba 277-8561, Japan}

\author{Y. Kinoshita}
\author{M. Tokunaga}
 \affiliation{Institute for Solid State Physics (ISSP), University of Tokyo, Chiba 277-8581, Japan}

\author{Y. Matsuda}
\affiliation{Department of Physics, Kyoto University, Kyoto 606-8502, Japan}

\author{K. Hashimoto}
\affiliation{Department of Advanced Materials Science, University of Tokyo, Kashiwa, Chiba 277-8561, Japan}\email{k.hashimoto@edu.k.u-tokyo.ac.jp}
 
\author{T. Shibauchi}%
 \affiliation{Department of Advanced Materials Science, University of Tokyo, Kashiwa, Chiba 277-8561, Japan}

\date{\today}

\begin{abstract}
We have developed a circularly polarized dielectric rutile (TiO$_2$) cavity with a high quality-factor that can generate circularly polarized microwaves from two orthogonal linearly polarized microwaves with a phase difference of $\pm\pi/2$ using a hybrid coupler.
Using this cavity, we have established a new methodology to measure the microwave Hall conductivity of a small single crystal of metal in the skin-depth region.
Based on the cavity perturbation technique, we have shown that all components of the surface impedance tensor can be extracted under the application of a magnetic field by comparing the right- and left-handed circularly polarized modes.
To verify the validity of the developed method, we performed test measurements on tiny Bi single crystals at low temperatures.
As a result, we have successfully obtained the surface impedance tensor components and confirmed that the characteristic field dependence of the ac Hall angle in the microwave region is consistent with the expectation from the dc transport measurements.
These results demonstrate a significant improvement in sensitivity compared to previous methods.
Thus, our developed technique allows more accurate microwave Hall measurements, opening the way for new approaches to explore novel topological quantum phenomena, such as time-reversal symmetry breaking in superconductors.
\end{abstract}

\maketitle

\section{\label{sec:1}INTRODUCTION}
The Hall effect, discovered by Edwin Hall in 1879\,\cite{Hall1879}, is one of the most fundamental phenomena in solid-state physics, in which electrons in a conductor placed in a magnetic field are subject to the Lorentz force, generating a transverse voltage perpendicular to the current flow in the conductor. The Hall conductivity, which is the off-diagonal component of the conductivity tensor, provides vital information on the type and concentration of charge carriers in a conductor. Hence, Hall measurements have long been performed on a wide variety of materials. Recent observations of the anomalous Hall effect in topological quantum materials with time-reversal symmetry breaking (TRSB) states, in which a Hall voltage is generated even in the absence of a magnetic field, have further highlighted the importance of Hall effect measurements\,\cite{Nagaosa2010}. Therefore, the Hall effect, which is usually investigated in the dc or low-frequency range, is one of the most important probes in solid-state physics.

In general, the Hall effect in the high-frequency region reflects the dynamical conductivity, which is defined as a complex number, allowing us to evaluate dissipative and non-dissipative information\,\cite{Arakawa2022}. Consequently, more information can be obtained compared to the Hall conductivity in the dc region, offering a more comprehensive insight into the microscopic properties of transport phenomena, especially when $\omega \tau$ cannot be ignored compared to 1 in, e.g., heavy fermion materials at low temperatures\,\cite{Shibauchi1997} or in electronic ordered states whose energy scale is close to the microwave region.

Such ac Hall conductivity measurements are usually carried out using the Faraday (Kerr) effect in the THz region\cite{Mittleman1997, Shimano2002, Ikebe2008, Ikebe2010, Shimano2011, Shimano2013, Okada2016, Okamura2020, Matsuda2020, Xia2006, Schemm2014}.
In these methods, the circular dichroism (CD), which is the response difference between the right and left-handed circularly polarized lights, rotates the polarization axis of the incident linearly polarized light, resulting in a transmitted (reflected) elliptical polarization when $\sigma_{xy}$ is finite, reflecting TRSB.
Since the rotation angle $\theta_{\rm{\,F(K)}}$ is proportional to the ac Hall conductivity $\sigma_{xy}$, one can obtain information on the dynamical conductivity from the rotation angle.
However, such magneto-optical measurements usually require a flat sample surface and a complex optical system.

So far, several measurements of the Hall effect in the microwave region have been reported\,\cite{Cooke1948, Portis1958, Hambleton1959, Watanabe1961, Nishina1961, Pethig1973, Sayed1975_1, Sayed1975_2, Fletcher1976, Ong1977, Cross1980, Ong1981, Eley1983, Kuchar1986, Dressel1991, Na1992, Chen1998, Prati2003, Al-Zoubi2005, Murthy2006, Murthy2008, Ogawa_bimodal2021, Ogawa_cuprate2021, ogawa_FeSe2023}.
In such measurements, the Hall component is detected using the two orthogonal magnetic field modes in a bimodal cavity.
However, most of the previous studies using this technique have been limited to measurements at room temperature on low-conductivity systems such as semiconductors\cite{Cooke1948, Portis1958, Hambleton1959, Watanabe1961, Nishina1961, Pethig1973, Sayed1975_1, Sayed1975_2, Fletcher1976, Ong1977, Cross1980, Ong1981, Eley1983, Kuchar1986, Dressel1991, Na1992, Chen1998, Prati2003, Al-Zoubi2005, Murthy2006, Murthy2008}.
Recently, Ogawa $et\,al$.\,\cite{Ogawa_bimodal2021} have enabled surface impedance tensor measurements in the skin depth region for materials with high conductivity at cryogenic temperatures by using both cross-shaped bimodal and ordinary cylindrical cavities.
This method is particularly useful for studying phenomena such as the flux flow Hall effect\,\cite{Ogawa_cuprate2021, ogawa_FeSe2023}. 
However, in this method, the geometric factor of the bimodal cavity depends on the sample and cavity geometries. Therefore, ordinary surface impedance measurements using a cylindrical cavity are required to obtain the Hall conductivity.

Recently, Arakawa $et\,al$. proposed a new method for the measurement of microwave Hall conductivity using a circularly polarized cavity\,\cite{Arakawa2022}. 
In this method, right- and left-handed circularly polarized eigenmodes are created using a cylindrical microwave cavity with a hybrid coupler\,\cite{Arakawa2022, Arakawa2019}. 
The differences in resonance frequency and $Q$-factor of each circularly polarized mode (i.e., CD) allow the determination of all components of the dynamical conductivity tensor by a non-contact method.
The previous study used circularly polarized electric field modes to measure the microwave Hall conductivity of high-resistance materials, such as a two-dimensional electron gas system.
In contrast, microwave impedance measurements of highly conductive materials generally require magnetic field modes\,\cite{Kitano2002}.
However, in the previous study\,\cite{Arakawa2022, Arakawa2019}, a copper cavity was used, resulting in a relatively low $Q$-factor ($Q<10000$) and low sensitivity, making precise Hall measurements for conductive materials with small Hall responses difficult.

To extend this technique to more metallic samples, we built a setup made up of a cylindrical cavity and equipped with a hybrid coupler to superpose two of the degenerate linear modes to produce the circular polarization. This setup houses a hollowed dielectric rutile made up of TiO$_2$ where the sample can be placed. This setup, which is named circularly polarized rutile cavity, is capable of attaining a high-$Q$.
In this paper, we present a new method to precisely measure the microwave Hall effect for tiny conductive samples.
We have developed a circularly polarized dielectric rutile \ce{TiO2} cavity with a higher $Q$-factor than the previous copper cavity even in high magnetic fields\,\cite{Arakawa2019, Arakawa2022}.
In addition, we have proposed a new protocol to measure all components of the surface impedance tensor in metals from the perturbation response to circularly polarized modes.
Furthermore, in this study, we have confirmed the validity of the developed method by performing test measurements on small bismuth (Bi) single crystals.
Our method enables more accurate measurements of the microwave Hall effect for conductive materials, which will allow us to explore topological quantum materials with an off-diagonal conductivity component, such as TRSB superconductors.

\begin{figure}[b]
    \centering
    \includegraphics[width=0.72\linewidth,]{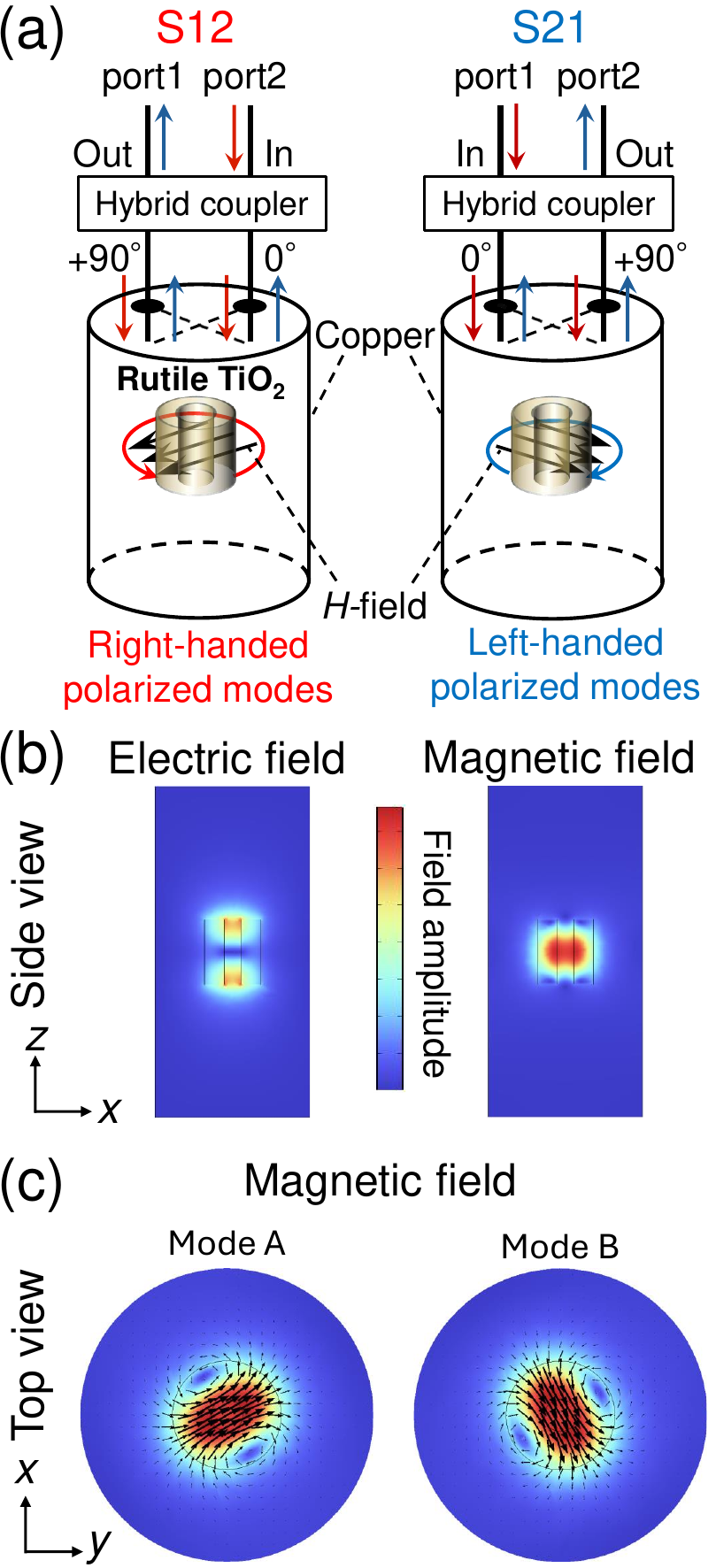}
    \caption{(a) Schematic of the developed circularly polarized dielectric rutile \ce{TiO2} cavity. The black arrows represent the ac (microwave) field generated inside the circularly polarized rutile cavity. The red (blue) arrows indicate the right (left)-handed polarized modes. The outer and inner diameters of the rutile is $\phi_{out}= 5.1$\,mm and $\phi_{in}= 1.5$\,mm, respectively, and the height is $L=5.9$\,mm. The rutile is placed inside an enclosed cylindrical copper wall (the diameter is $\phi = 14$\,mm and the height is $L=30$\,mm). Here, the $z$ direction of the cylindrical axis for the rutile cavity corresponds to the $c$-axis direction of the rutile, and the $xy$ plane perpendicular to it corresponds to the $ab$ plane. [(b) and (c)] Simulations of electromagnetic field distribution for the hybrid mode. (b) The magnitude distributions of the electric and magnetic fields of our focused mode in the vertical cross-section of the cavity. (c) The magnetic field vector distributions of the hybrid mode in (b) at the horizontal cross section at the middle of the rutile, which are decomposed into two orthogonal and degenerate linearly polarized modes A and B.}
\label{fig:circulaly_cav}
\end{figure}

\section{\label{sec:2}EXPERIMENTAL METHOD}
\subsection{\label{sec:2-A} Circularly polarized microwave cavity}
Circular polarization is the superposition of two orthogonal and degenerate linear polarized waves with a phase difference of $\pm\frac{\pi}{2}$.
To realize circularly polarized eigenmodes inside a microwave cavity, an apparatus that can introduce a relative phase difference between two orthogonal and degenerate linearly polarized modes is required. 
To achieve this, we adopted a cylindrical circularly polarized cavity equipped with a hybrid coupler, as proposed by Arakawa $et\,al$.\,\cite{Arakawa2019, Arakawa2022}
The main advantage of this system is that the right- and left-handed modes of circular polarization can be easily controlled by switching the input port, which reverses the $\pm \frac{\pi}{2}$ phase difference.
In other words, the right\,($+$)- and left\,($-$)\,-handed circular polarization modes can be controlled by the vector network analyzer through switching of the scattering parameter ($S$-parameter) measurements between $S_{12}$ and $S_{21}$.
Furthermore, the hybrid coupler has a compact structure, allowing easy installation in a cryostat.

Microwave cavities with high $Q$ factors are required to overcome the measurement limits on samples with large CD responses\,\cite{Arakawa2022, Arakawa2019} and materials with small Hall responses. This can be achieved using superconducting microwave cavities, such as Pb and Al cavities. However, the type-I superconductors cannot maintain a high-$Q$ in magnetic fields due to their small critical fields. 
In contrast, type-II superconducting cavities have high $Q$ values ($>10^5$) even in high magnetic fields below their upper critical fields. However, their performance degrades significantly above their critical temperatures\,\cite{Alesini2019, Golm2022, Posen2023}.
Thus, we have employed a dielectric rutile \ce{TiO2} cavity with a high $Q$-factor even at high temperatures and in high magnetic fields.
Our rutile \ce{TiO2} cavity can maintain a high-$Q$ value of $10^5$ even at high temperatures up to $\sim$100\,K and in higher magnetic fields, as shown in Fig.\ref{fig:Q(T)}.
Previous studies have reported that the rutile cavity can maintain a high $Q$ factor of $\sim 10^6$ with a frequency drift of the cavity less than 1 Hz/h even in high magnetic fields\,\cite{Huttema2006}.

Figure\,\ref{fig:circulaly_cav}(a) shows an overview of the developed circularly polarized rutile cavity.
A cylindrical rutile \ce{TiO2} is placed at the center of a cylindrical copper enclosure, on which the input and output ports are equipped through a hybrid coupler. Figures\,\ref{fig:circulaly_cav}(b) and (c) show the electromagnetic field distributions of a hybrid mode in the circularly polarized rutile cavity used in this study obtained from the electromagnetic field simulations (software: COMSOL).
In general, the introduction of a dielectric material in the cavity leads to complex hybrid modes, in which both the electric and magnetic fields vary in the axial and angular directions\,\cite{Zaki1983,Zaki1988}. The mode shown in Fig.\,\ref{fig:circulaly_cav}(b) has a similar magnetic field distribution to the TM$_{110}$ mode, but it is a hybrid mode with axial distributions for both the electric and magnetic fields.
The magnetic field distribution is concentrated at the center of the rutile (Fig.\,\ref{fig:circulaly_cav}(b), right), while the electric field is weak (Fig.\,\ref{fig:circulaly_cav}(b), left).
The magnetic field vector distributions of the hybrid mode in Fig.\,\ref{fig:circulaly_cav}(b) can be decomposed into two orthogonal and degenerate linearly polarized modes, as shown in Fig.\,\ref{fig:circulaly_cav}(c).
Our electromagnetic field simulations show that the central rutile confines almost all the microwave fields, significantly reducing the shielding currents in the surrounding copper walls, which results in a high $Q$-factor.
Furthermore, using a small rutile cavity, the ratio of the sample volume to the electromagnetic field volume (filling factor) increases, resulting in an increase in sensitivity.
In our setup, the resonance frequency obtained from the electromagnetic field simulation is $f_0 \sim 4.24$\,GHz at 4 K, while the actual experimental value for the circularly polarized rutile cavity without a sample is $f_0 \sim 4.17$\,GHz. The experimentally obtained $Q$-factor without a sample is $Q_0 \sim 1\times10^5$ at 10\,K, which is lower than the previously reported $Q_0 \sim 1\times10^6$\,\cite{Huttema2006}. This is due to the small diameter of the copper enclosure used here, which causes energy loss due to leakage of electromagnetic fields around the copper walls. 
However, the obtained $Q$-factor is significantly higher than those in the previously reported circularly polarized\,\cite{Arakawa2022} and bimodal cavities\,\cite{Ogawa_bimodal2021}, allowing for more sensitive microwave Hall effect measurements in a magnetic field, as shown in the following sections.
\subsection{\label{sec:2-B} Surface impedance tensor measurements}
Surface impedance is a quantity that represents the resistance per unit area to the ac fields in the skin depth region, and the surface impedance tensor in the $xy$-plane is expressed as follows\,\cite{Ogawa_bimodal2021}:
\begin{equation}
\label{Z_tensor}
    \hat{Z} = 
    \begin{pmatrix}
        Z_{\rm{L}}, & Z_{\rm{H}} \\
        -Z_{\rm{H}}, & Z_{\rm{L}}
    \end{pmatrix}
    =
    \begin{pmatrix}
    R_{\rm{L}} + iX_{\rm{L}}, & R_{\rm{H}} + iX_{\rm{H}} \\
    -R_{\rm{H}} - iX_{\rm{H}}, & R_{\rm{L}} + iX_{\rm{L}}
    \end{pmatrix},
\end{equation}
where $Z_{\rm{L(H)}}$ is the diagonal (off-diagonal) component of the surface impedance tensor, and $R_{\rm{L(H)}}$ and $X_{\rm{L(H)}}$ are the real and imaginary parts of the diagonal (off-diagonal) component, which are the surface (surface Hall) resistance and surface (surface Hall) reactance, respectively.

\begin{figure}[b]
\centering
\includegraphics[width=0.8\linewidth]{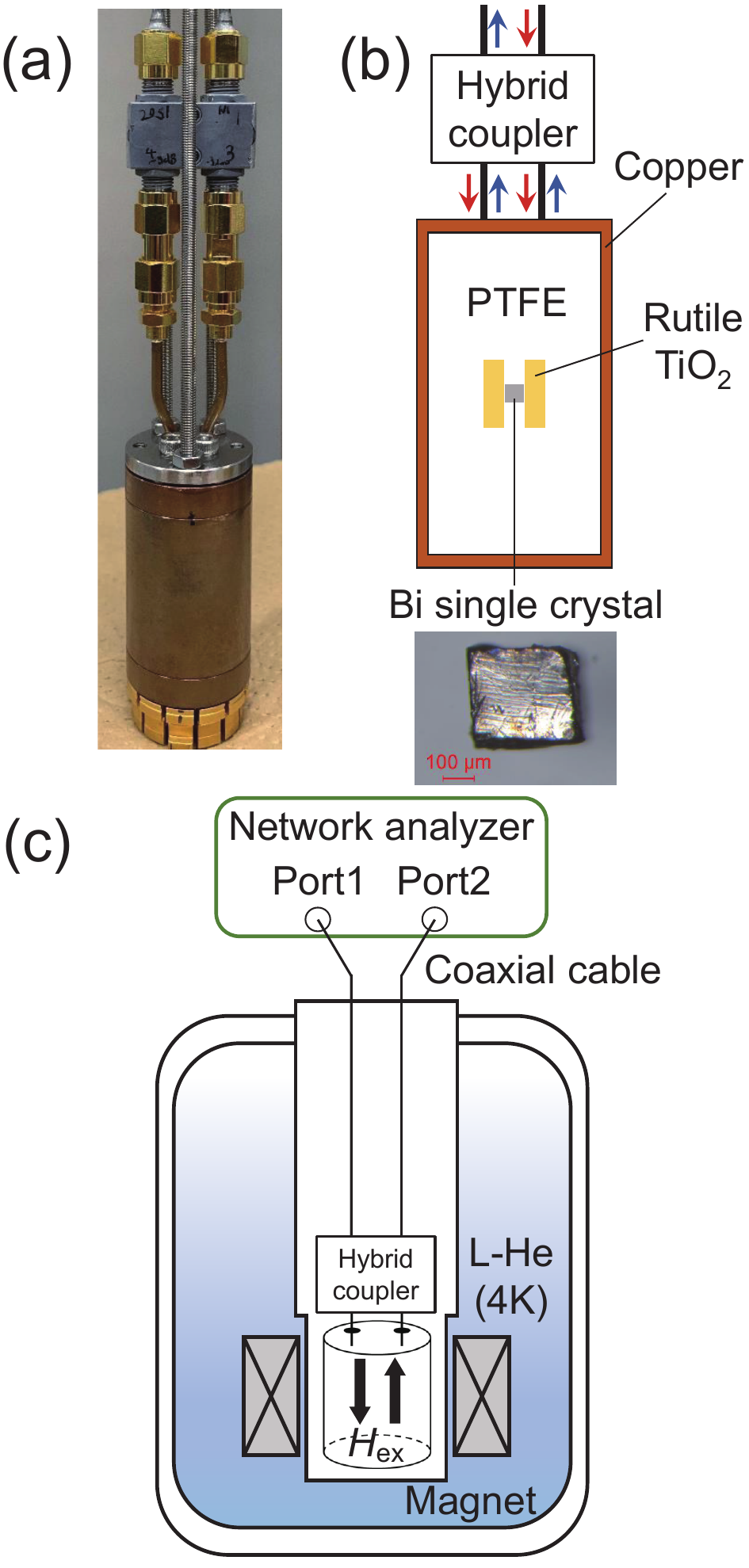}
\caption{\label{fig:setup} Experimental setup for the surface impedance tensor measurements using the circularly polarized rutile cavity. (a) Photograph of the cavity fabricated in this study. The top of the cavity is attached to the input and output coaxial cables via a hybrid coupler. A PPMS pack is attached to the bottom of the cavity to achieve good thermal contact with the sample space of the PPMS. (b) Schematic of the circularly polarized rutile cavity. A rutile cavity with a hollowed center is placed inside a copper cavity. In this study, a single-crystal sample of Bi was placed in the center of the rutile cavity. (c) Schematic diagram of the entire measurement system. The temperature and external dc magnetic field (black arrows) are controlled by the PPMS control system. The vector network analyzer (Keysight P5027A) located at room temperature measures $S_{12}$ and $S_{21}$ spectra through the semi-rigid coaxial cables.}
\end{figure}

Most of the surface impedance measurements using the cavity perturbation technique reported so far\,\cite{Huttema2006, Donovan1993, Shibauchi1994, Hashimoto2009, Hashimoto2012} utilize the magnetic field of the linearly polarized TE$_{011}$ mode. This method allows the detection of only the diagonal component of the surface impedance, $Z_{\rm{L}}$, from the changes in resonance frequency $f$ and $Q$-factor.
To the best of our knowledge, surface impedance measurements using magnetic fields of circularly polarized modes have not been performed so far. Therefore, in this study, we derive the relationship between the resonance characteristics of magnetic fields of  circularly polarized modes and the surface impedance tensor in the cavity perturbation technique.
As a result, we find the following relationship (the detailed derivation is described in Appendix\,\ref{AP:Derivation}):
\begin{equation}
\label{resonance_and_Z}
    \Delta\Bigl(\frac{1}{2 Q_{\pm}}\Bigr) - i\Bigl(\frac{\Delta f_{\pm}}{f_0} + C\Bigr)
    = G (Z_{\rm{L}} \mp i Z_{\rm{H}}),
\end{equation}
where $C$ is the metallic shift, and $G$ is the geometric factor. 
The indices $\pm$ and 0 represent the right\,(+) and left\,(-)-handed polarized modes and the blank without a sample, respectively, and $\Delta$ represents the difference before and after sample insertion. 
From Eq.\,(\ref{resonance_and_Z}), the components of the surface impedance tensor can be given by using the resonant frequency $f$ and the $Q$-factor as follows:
\begin{subequations}
\begin{align}
&\frac{1}{2 Q_{+}} - \frac{1}{2 Q_{0}} = G (R_{\rm{L}} + X_{\rm{H}}), \label{Q_plus_RL_XH}
\\
&\frac{1}{2 Q_{-}} - \frac{1}{2 Q_{0}} = G (R_{\rm{L}} - X_{\rm{H}}), \label{Q_minus_RL_XH}
\\
& -\frac{f_{+} - f_0}{f_0} = G (X_{\rm{L}} - R_{\rm{H}}) + C,\label{f_plus_RH_XL}
\\
& -\frac{f_{-} - f_0}{f_0} = G (X_{\rm{L}} + R_{\rm{H}}) + C, \label{f_minus_RH_XL}
\end{align}
\end{subequations}
where $f_0$ and $Q_0$ are the resonance frequency and $Q$-factor without a sample, respectively.
From Eqs.\,(\ref{Q_plus_RL_XH})-(\ref{f_minus_RH_XL}), each component of the surface impedance tensor is expressed as
\begin{subequations}
\begin{align}
&R_{\rm{L}} = \frac{1}{4G} \left( \frac{1}{Q_+}+\frac{1}{Q_-} - \frac{2}{Q_0} \right), \label{RL_fpm}
\\
& X_{\rm{L}} = -\frac{1}{2G} \left( \frac{f_+ + f_- - 2f_0}{f_0} +2C  \right), \label{XL_Qpm}\\
&R_{\rm{H}} = \frac{1}{2G} \left( \frac{f_+ - f_-}{f_0} \right), \label{RH_fpm}
\\
&X_{\rm{H}} = \frac{1}{4G} \left( \frac{1}{Q_+} - \frac{1}{Q_-} \right). \label{XH_Qpm}
\end{align}
\end{subequations}
Here we emphasize that Eqs.\,(\ref{RH_fpm}) and (\ref{XH_Qpm}) indicate that the difference between the resonance frequencies $f_{\pm}$ and the $Q$-factors $Q_{\pm}$ in the right and left-handed modes, i.e., CD, is proportional to the off-diagonal term of the surface impedance tensor.
Our present method allows in situ measurements of right- and left-handed modes in the same cavity, and $G$ and $C$ are uniquely determined (the detailed analysis procedures are described in the next section).
Thus, compared to the previous method\,\cite{Ogawa_bimodal2021}, which requires multiple resonators and several assumptions, this method can more easily and accurately obtain the surface impedance tensor.

\begin{figure}[tb]
\centering
\hspace*{-1.5cm}
\includegraphics[width=0.85\linewidth]{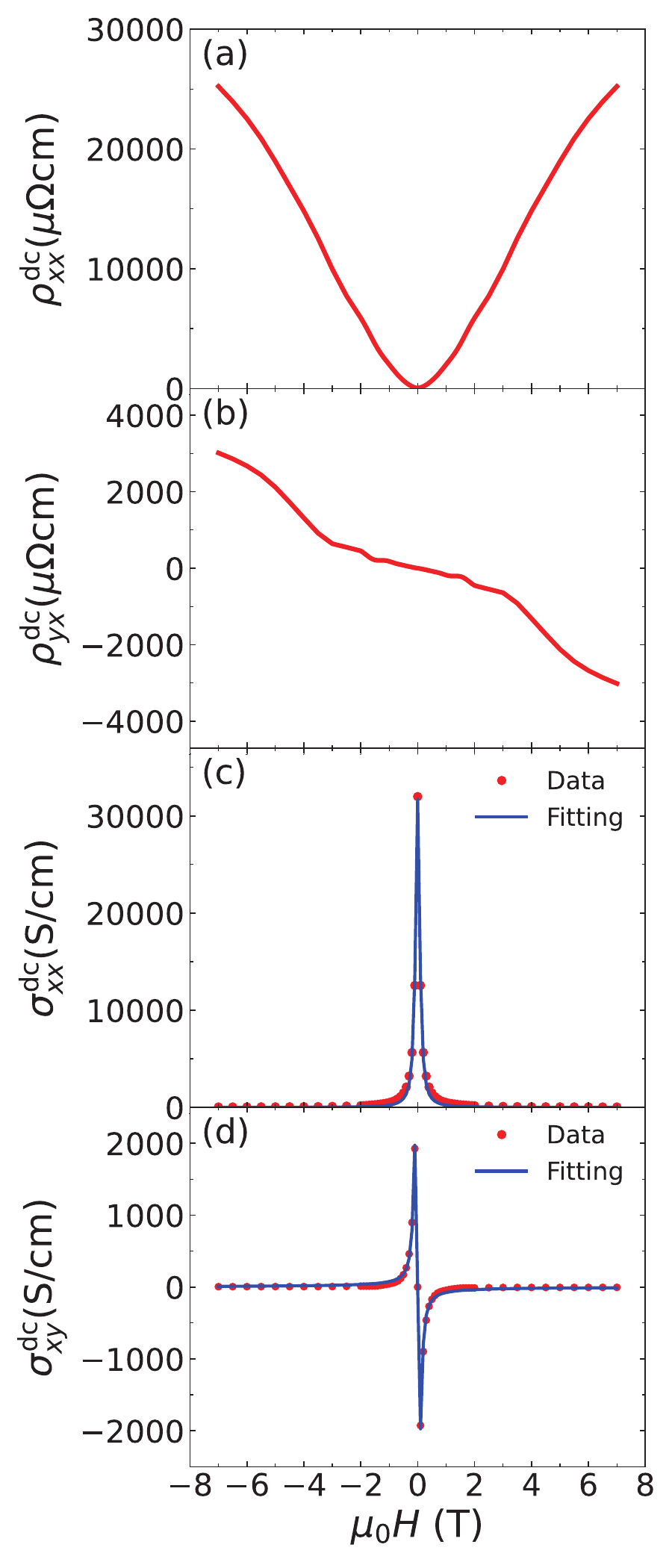}
\caption{[(a) and (b)] Magnetic field dependence of the longitudinal resistivity $\rho_{xx}$ (a) and the Hall resistivity $\rho_{yx}$ (b). [(c) and (d)] Magnetic field dependence of the longitudinal conductivity $\sigma_{xx}$ (c) and Hall conductivity $\sigma_{xy}$ (d) extracted from Eqs.\,(\ref{sigma_xx_invert}) and (\ref{sigma_xy_invert}). The red circles represent the experimental data, and the blue solid lines represent the fitting results.}
\label{fig:DC}
\end{figure}
\begin{figure*}[tb]
    \centering
    \includegraphics[width=\linewidth]{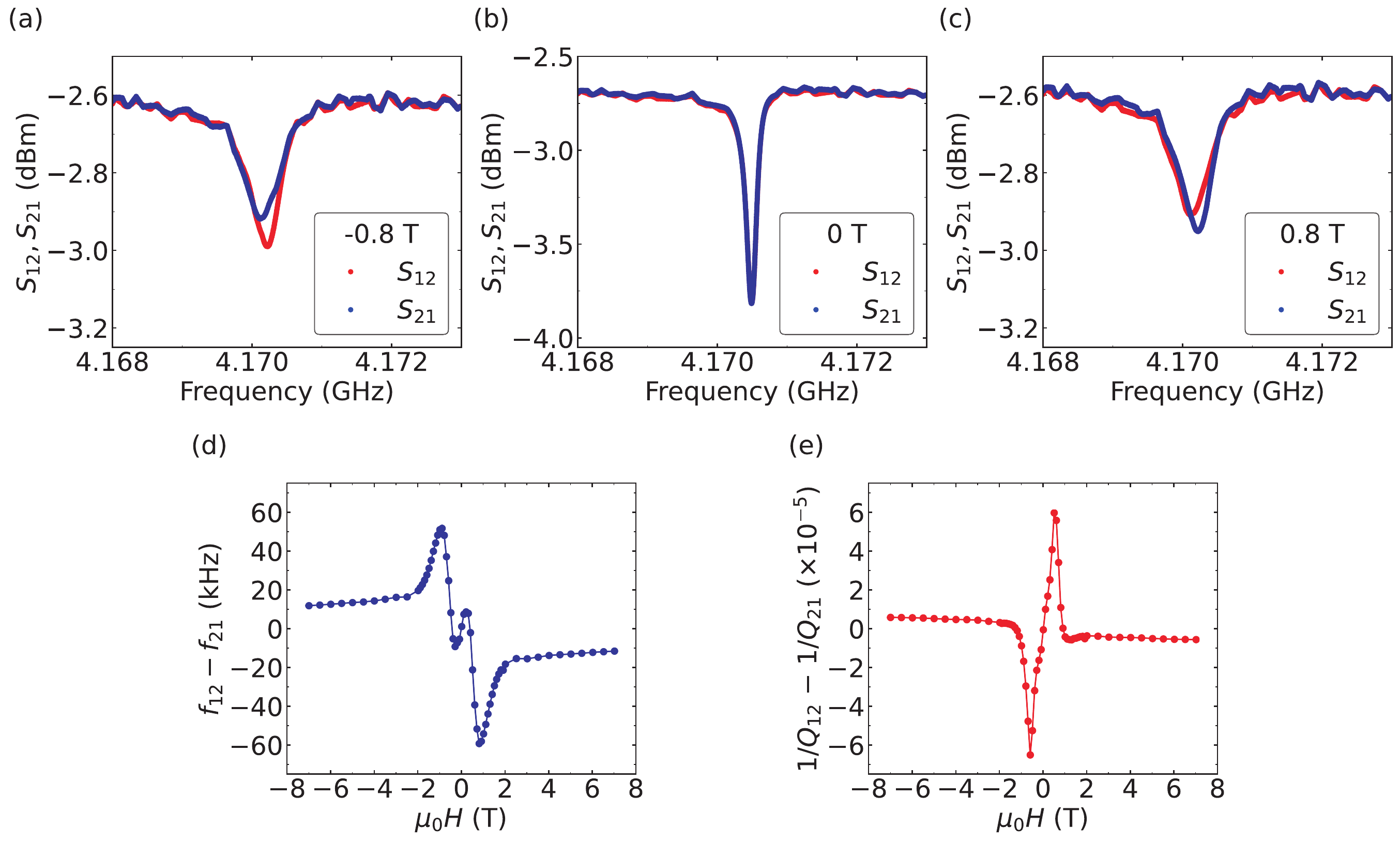}
    \caption{[(a)-(c)] Resonance spectra of $S_{12}$ (red circles) and $S_{21}$ (blue circles) measured at 10\,K for the Bi single crystal at external fields of $-0.8$\,T (a), 0\,T (b), and $+0.8$\,T (c). [(d) and (e)] Difference in resonant frequency (d) and 1/$Q$ (e) between $S_{12}$ and $S_{21}$ at 10 K as a function of magnetic field.}
\label{fig:spectra}
\end{figure*}

Finally, we derive the relationship between the surface impedance tensor and the ac Hall conductivity.
The conductivity tensor $\tilde \sigma$ and the surface impedance tensor are related by the following expression\,\cite{Ogawa_bimodal2021}:
\begin{equation}
\label{Z_sigma}
   \tilde Z = \dfrac{1}{\delta} {\tilde \sigma}^{-1},
\end{equation}
where $\delta$ denotes the skin depth.
From this relation, $Z_{\rm{L}}$ and $Z_{\rm{H}}$ are represented by
\begin{subequations}
\begin{eqnarray}
Z_{\rm{L}} = \sqrt{\dfrac{i\mu_0\omega}{\sigma(\omega)}} \dfrac{1}{\sqrt{1 \mp i \tan \theta}}, \label{ZL_sigma}
\\
Z_{\rm{H}} = \sqrt{\dfrac{i\mu_0\omega}{\sigma(\omega)}} \dfrac{\tan \theta}{\sqrt{1 \mp i \tan \theta}}, \label{ZH_sigma}
\end{eqnarray}
\end{subequations}
where $\mu_0$ is the vacuum permeability, $\omega = 2\pi f$ is the measurement angular frequency, and  $\sigma(\omega) = ne^2\tau / m^{\ast}(1-i\omega)$ is the ac conductivity (here $e$ is the elementary charge, $n$ is the carrier density, $m^{\ast}$ is the effective mass of charge carriers, and $\tau$ is the relaxation time).
Note that $\theta$ denotes the Hall angle ($\tan\theta=\sigma_{xy}/\sigma_{xx}$). 
From Eqs.\,(\ref{ZL_sigma}) and (\ref{ZH_sigma}), $\theta$ can be related to the ratio of $Z_{\rm{H}}$ to $Z_{\rm{L}}$ by the following expression:
\begin{equation}
  \label{tan_Z_H Z_L}
   \tan\theta = \frac{Z_{\rm{H}}}{Z_{\rm{L}}}.
\end{equation}
Therefore, the ac Hall angle can be determined by measuring the surface impedance tensor, providing information on the ac Hall conductivity in the skin depth region.

\subsection{\label{sec:2-C} Experimental setup and analysis procedure}
Our experimental setup for the measurement of the surface impedance tensor is shown in Fig.\ref{fig:setup}.
The rutile cavity equipped with a hybrid coupler (Orient Microwave, BL32-6347-00), is inserted into a commercial cryostat (Physical Property Measurement System (PPMS), Quantum Design) using a homemade probe.
The transmission spectra of $S_{12}$ and $S_{21}$ are measured by a vector network analyzer (Keysight P5027A) in the room temperature part through semi-rigid coaxial cables.
The cylindrical rutile \ce{TiO2} grown by the Verneuil method (Crystal Base Co., Ltd.) is positioned at the center of the copper enclosure with polytetrafluoroethylene (PTFE) spacers.
A single crystal of Bi (320\,$\mu$m (binary)$\times 290$\,$\mu$m (bisectrix)$\times 65$\,$\mu$m (trigonal)) synthesized by the Czochralski method\,\cite{Hiruma1982} is placed at the center of the cylindrical rutile.
A sample pack of PPMS was attached to the bottom of the cavity for thermal contact. 
External dc magnetic fields ($-7$\,T $\leq \mu_0 H \leq 7$\,T) were applied from negative to positive in a direction perpendicular to the Bi single-crystal sample plane (parallel to the triangular axis).
Our analysis procedure in the surface impedance tensor measurements is as follows:\\
(1)\,Determination of $R_{\rm{L}}$ and $X_{\rm{L}}$: In Eq.\,(\ref{ZL_sigma}), when the Hagen-Rubens limit $\omega\tau\ll 1$ is satisfied, $Z_{\rm{L}}$ can be replaced by
\begin{equation}
  \label{ZL_HRlimit}
   Z_{\rm{L}}= (1+i)\sqrt{\dfrac{\mu_0 \omega \rho^{\rm{dc}}_{xx}}{2}} \dfrac{1}{\sqrt{1 \mp i \tan\theta}},
\end{equation}
where $\rho^{\rm{dc}}_{xx}$ is the dc longitudinal resistivity.
From this equation, the real and imaginary parts $R_{\rm{L}}$ and $X_{\rm{L}}$ are given by
\begin{subequations}
\begin{eqnarray}
R_{\rm{L}} = \text{Re} \Bigl[ (1+i)\sqrt{\dfrac{\mu_0 \omega \rho^{\rm{dc}}_{xx}}{2}} \dfrac{1}{\sqrt{1 \mp i \tan\theta}} \Bigr], \label{RL_HRlimit}
\\
X_{\rm{L}} = \text{Im} \Bigl[(1+i)\sqrt{\dfrac{\mu_0 \omega \rho^{\rm{dc}}_{xx}}{2}} \dfrac{1}{\sqrt{1 \mp i \tan\theta}} \Bigr]. \label{XL_HRlimit}
\end{eqnarray}
\end{subequations}
As shown in the next section, since $\omega\tau \ll 1$ is well satisfied in our Bi single crystal, we can obtain $R_{\rm{L}}$ and $X_{\rm{L}}$ from Eqs.\, (\ref{RL_HRlimit}) and (\ref{XL_HRlimit}). Here, the correct value of $\tan\theta$ is unknown in step (1) since $\tan\theta$ is a quantity to be finally determined. Therefore, in step (1), we substitute a certain value $\theta_0$ for the Hall angle with reference to the Hall angle obtained from the dc measurements. Then, using $R_{\rm L}$ and $X_{\rm L}$ obtained in step (1), $\tan\theta$ is determined through steps (2) and (3). If the value of $\tan\theta$ obtained in step (3) coincides with $\tan\theta_0$ in step (1), the Hall angle can be obtained self-consistently.\\
(2)\,Determination of $G$, $C$, $R_{\rm{H}}$, and $X_{\rm{H}}$: First, we measure the $S_{12}$ and $S_{21}$ spectra corresponding to the right- and left-handed modes as a function of magnetic field by switching the input and output ports of the circular polarization rutile cavity, from which we obtain $f_{\pm}(H)$ and $Q_{\pm}(H)$. Then, we calculate $G$ and $C$ by comparing Eqs.\,(\ref{RL_fpm}) and (\ref{XL_Qpm}) with Eqs.\,(\ref{RL_HRlimit}) and (\ref{XL_HRlimit}). After that, we extract $R_{\rm{H}}$ and $X_{\rm{H}}$ from Eqs.\,(\ref{RH_fpm}) and (\ref{XH_Qpm}) by using $G$.\\
(3)\,Calculation of $\tan\theta$: Using each component of the obtained surface impedance tensor, we calculate the ac Hall angle using the following equation:
\begin{equation}
   |\tan \theta| = \dfrac{|Z_{\rm{H}}|}{|Z_{\rm{L}}|} = \dfrac{\sqrt{R_{\rm{H}}^2 + X_{\rm{H}}^2}}{\sqrt{R_{\rm{L}}^2 + X_{\rm{L}}^2}}. \label{tan_from_Z}
\end{equation}
Finally, to confirm the validity of the present measurements and analysis, we compare the results of $\tan\theta$ obtained from Eq.\,(\ref{tan_from_Z}) with those obtained from the dc transport measurements. The sign of $\tan\theta$ in Eq.\,(\ref{tan_from_Z}) was determined based on the results of $\tan\theta$ from the dc measurements (see Appendix\,\ref{AP_DC} for details).

\section{\label{sec:3}EXPERIMENTAL RESULTS}
\subsection{\label{sec:3-A} DC resistivity measurements}
As shown in Secs.\,\ref{sec:2-B} and \ref{sec:2-C}, since the dc resistivity is necessary to obtain the surface impedance tensor, we performed dc resistivity measurements on a Bi single crystal.
In this section, we present the results of the dc resistivity measurements (see Appendix\,\ref{AP_DC} for more details).
Figures\,\ref{fig:DC}(a) and (b) show the magnetic field dependence of dc longitudinal resistivity ($\rho^{\rm{dc}}_{xx}$) and Hall resistivity ($\rho^{\rm{dc}}_{yx}$) in our Bi single crystal at 10\,K, respectively, where the $x$ and $y$ directions correspond to the binary and bisectrix axes.
$\rho^{\rm{dc}}_{xx}$ shows a giant magnetoresistance due to the high mobility and carrier compensation in Bi (Fig.\,\ref {fig:DC}(a)), as reported in previous works\,\cite{Fauque2009}. 
We have also observed a large $\rho^{\rm{dc}}_{yx}$ in high magnetic fields (Fig.\,\ref{fig:DC}(b)), reflecting the lower carrier density compared to other conventional metals.
We extracted the longitudinal and transverse (Hall) conductivity $\sigma^{\rm{dc}}_{xx}$ and $\sigma^{\rm{dc}}_{xy}$ from $\rho^{\rm{dc}}_{xx}$ and $\rho^{\rm{dc}}_{yx}$ and analyzed them by simultaneously fitting with the multi-band model (see Appendix C for more details).
As a result, we obtained the electron (hole) carrier density $n_{\rm{e(h)}} = 8.7 \times 10^{17}(8.3 \times 10^{17})$\,cm$^{-3}$ and the mobility $\mu_{\rm{e(h)}} = 1.2 \times 10^{5}(9.2\times 10^{4})$\,cm$^{2}$V$^{-1}$s$^{-1}$, respectively.
These results are consistent with the previous studies\cite{Michenaud1972}.
From $\mu_{\rm{e(h)}}=e\tau/m^*$ with $m^*=0.051m_0$\cite{Everett1962}, we obtained $\tau=3.4 (2.7)\times10^{-12}$\,s.
In the present cavity measurements, $\omega=2\pi f \sim 2.6\times10^{10}$\,Hz, and thus the Hagen-Rubens limit $\omega\tau \ll 1$ is satisfied in our Bi single crystal.

\begin{figure}[t]
\hspace*{-1.5cm}
\includegraphics[width=0.8\linewidth, clip]{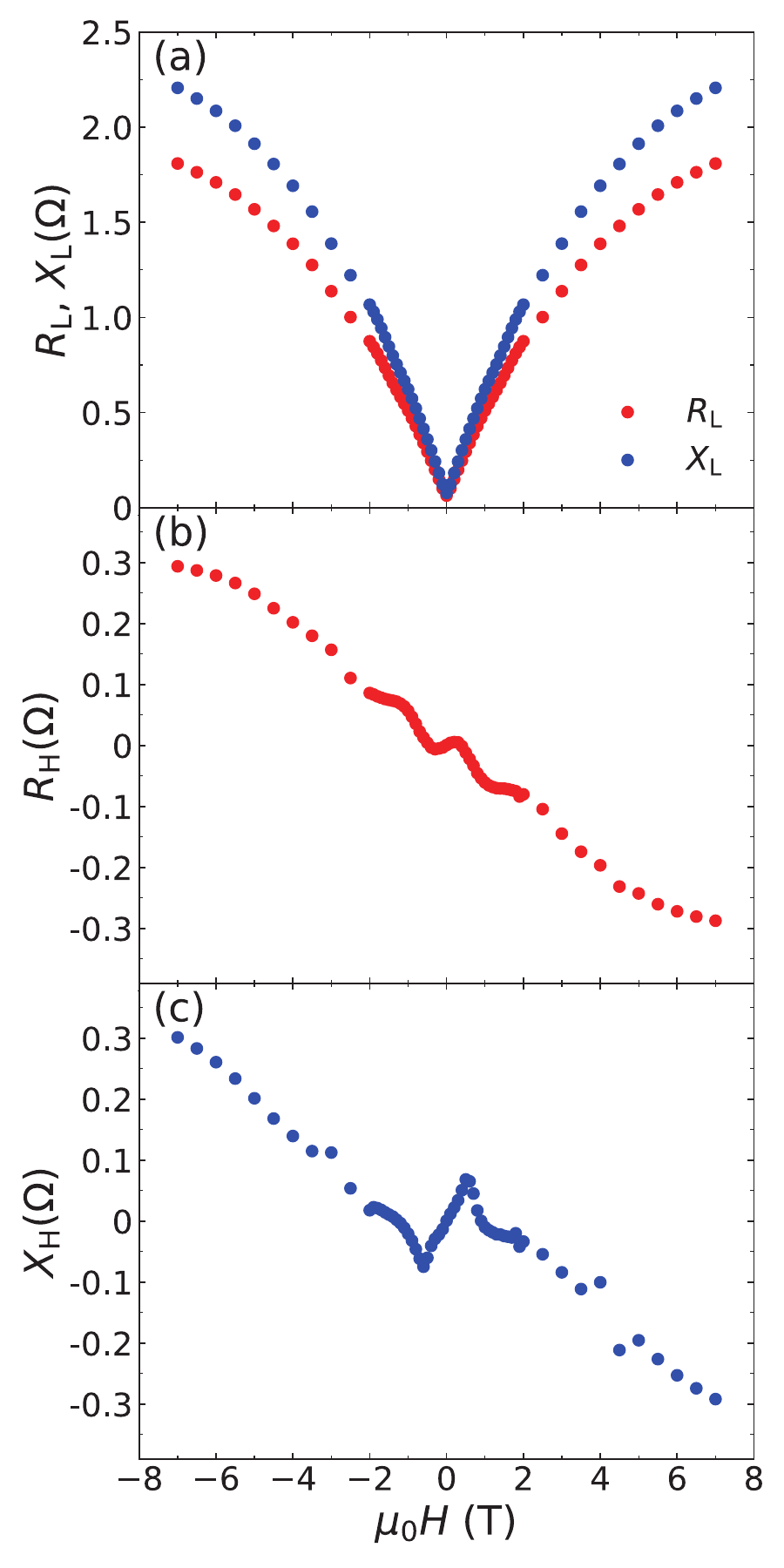}
\caption{Magnetic field dependence of each component
of the surface impedance tensor measured at 10\,K. [(a)-(c)] The real part $R_{\rm{L}}$ and imaginary part $X_{\rm{L}}$ of the diagonal components (a), and the real part $R_{\rm{H}}$ (b) and the imaginary part $X_{\rm{H}}$ (c) of the off-diagonal components are shown.}
\label{fig:impedance}
\end{figure}

\subsection{\label{sec:3-B} Surface impedance tensor measurements}
This section presents the results of surface impedance tensor measurements using our circularly polarized rutile cavity.
Figures\,\ref{fig:spectra}(a)-(c) show the resonace spectra of $S$-parameters for two polarization modes $S_{12}$ and $S_{21}$ measured at $-0.8$\,T, 0\,T, and $+0.8$\,T at 10\,K for the Bi single crystal, respectively.
At zero magnetic field, $S_{12}$ and $S_{21}$ show no splitting (Fig.\,\ref{fig:spectra}(b)). In sharp contrast, $S_{12}$ and $S_{21}$ show clear splitting in the finite external magnetic fields (Figs.\,\ref{fig:spectra}(a) and (c)).
Figures\,\ref{fig:spectra}(d) and (e) show the difference in the resonance frequency, $f_{12}-f_{21}$, and the energy loss, $1/Q_{12}-1/Q_{21}$, between the right- and left-handed modes obtained from the fitting of the $S_{12}$ and $S_{21}$ spectra for each magnetic field, where the indices of 12 and 21 correspond to the right ($+$)- and left ($-$)-handed modes.
We observed distinct field-induced differences between the right- and left-handed modes, which are symmetric with respect to the spectrum at zero field.
The origin of the dip structure at the low magnetic field will be discussed later.
Following the procedure described in Sec.\,\ref{sec:2-C}, we obtained each component of the surface impedance tensor from the obtained $f_{12}$, $f_{21}$, $Q_{12}$, and $Q_{21}$, as shown in Figs.\,\ref{fig:spectra}(d) and (e).

\begin{figure}[t]
\hspace*{-1.0cm}
\includegraphics[width=0.9\linewidth, clip]{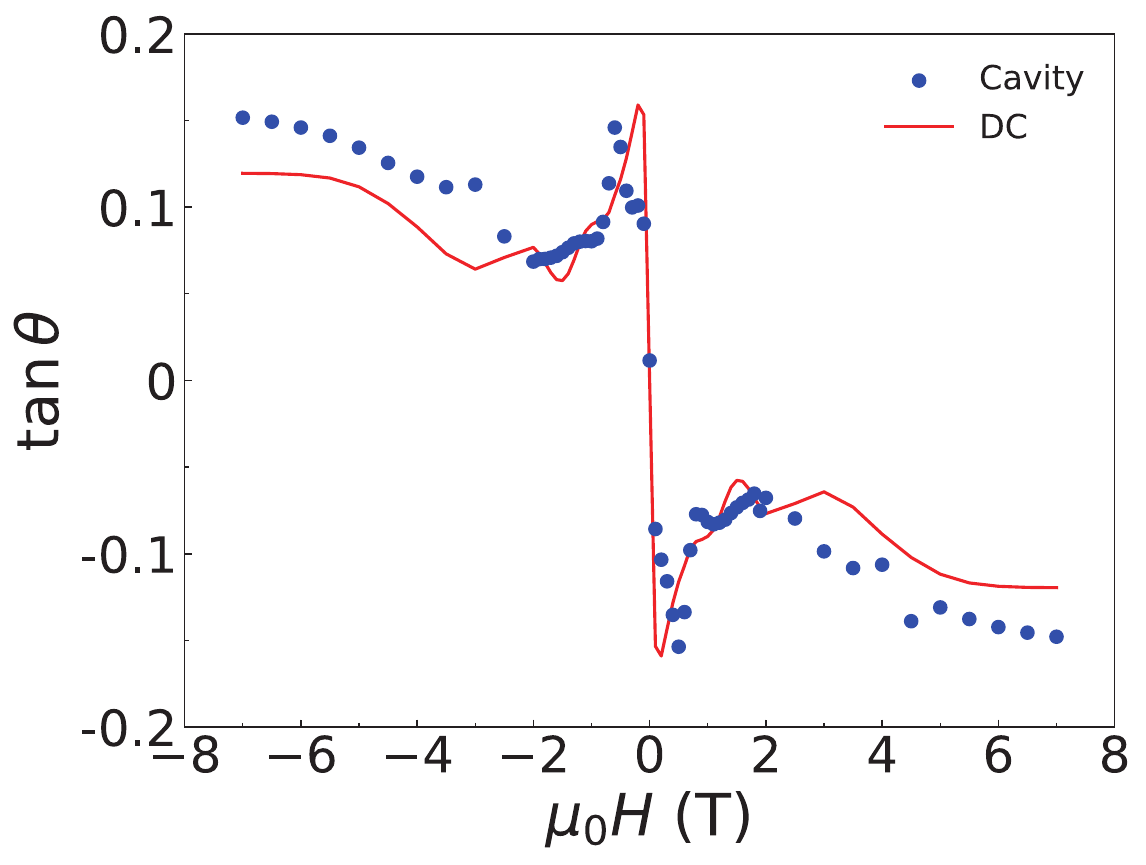}
\caption{Magnetic field dependence of $\tan\theta$ obtained from the microwave cavity measurements (blue circles) and dc transport measurements (red solid line) at 10 K.}
\label{fig:tan}
\end{figure}

First, we show the magnetic field dependence of $R_{\rm{L}}$ and $X_{\rm{L}}$ estimated from Eqs.\,(\ref{RL_HRlimit}) and (\ref{XL_HRlimit}) in Fig.\,\ref{fig:impedance}(a).
$R_{\rm{L}}$ and $X_{\rm{L}}$ show large magnetic field dependence, reflecting the large magnetoresistance of Bi. 
Next, we derive $R_{\rm{H}}$ and $X_{\rm{H}}$ as a function of magnetic field, as shown in Figs.\,\ref{fig:impedance}(b) and (c), respectively, which are obtained from Eqs.\,(\ref{RH_fpm}) and (\ref{XH_Qpm}).
The wiggle structures at low magnetic fields ($\sim 0.2\,$T), which correspond to the dip structures seen in Figs.\,\ref{fig:spectra}(d) and \ref{fig:spectra}(e), may reflect the circular dichroism for the cyclotron resonances of the hole carriers in Bi\,\cite{Visser16}.
Finally, we obtain the Hall angle for each magnetic field from Eq.\,(\ref{tan_from_Z}) as shown in Fig.\,\ref{fig:tan}.
The field dependence of $\tan\theta$ obtained from the circularly polarized rutile cavity is in good agreement with the dc results.
Here, we emphasize that $\tan\theta$ obtained from the cavity measurements is the ac Hall angle in the microwave region, and it is expressed $\tan \theta = \omega_{c}\tau/(1-i\omega\tau)$ in the Drude model, where $\omega_c=eB/m^{\ast}$ is the cyclotron angular frequency (here $B$ is the magnetic flux density).
As described in the previous section, since our Bi single crystal satisfies the condition $\omega\tau \ll 1$, the finite frequency correction term $1/(1-i\omega\tau)$ can be negligible, resulting in the coincidence of the dc and ac Hall angles.
This result confirms the validity of our developed method.

\section{\label{sec:4}SUMMARY}
In conclusion, we have developed a circularly polarized microwave cavity with rutile \ce{TiO2} that can maintain a high $Q$-factor even in magnetic fields and constructed a method to measure the surface impedance tensor. 
From measurements of tiny Bi single crystals, we have demonstrated that all components of the surface impedance tensor can be accessed from the response of the circularly polarized magnetic field modes.
The advantages of the developed method are as follows:
(1) Utilizing the cavity perturbation technique, it enables highly sensitive measurements on tiny single crystals in the microwave frequency range.
(2) This non-contact method is particularly useful for samples and systems that are difficult to measure by the usual contact method.
(3) The experimental setup is straightforward, allowing easy switching between the right- and left-handed circular polarization at cryogenic temperatures.
These advantages can allow precise measurements of the Hall effect in the microwave region, which are particularly useful for systems such as superconductors, where the dc Hall resistance, as well as the dc resistance, becomes zero.
The Hall effect measurements in superconductors in the high-frequency regime have the potential to probe TRSB in the superconducting state\,\cite{Xia2006, Schemm2014}.
Therefore, our newly developed technique provides new insights into Hall measurements in the finite frequency range, offering crucial understanding of exotic phenomena associated with TRSB states in condensed matter physics. 

\begin{acknowledgments}
We thank M. Kondo, A. Yamada, K. Kato, and Y. Onishi for fruitful discussions, and K. Matsuura, Y. Nakamura, and Y. Uwatoko for technical support, and N. Miura for sample preparation. This work was supported by CREST (JPMJCR19T5) from Japan Science and Technology (JST), Grants-in-Aid for Scientific Research (KAKENHI) (Nos.\,JP23H00089, JP23H04862, JP22J21896, JP22H00105, JP22K18681, JP22K18683, JP22K20349, JP22H01964, JP22H01936, JP22K03522, JP21H01793, and JP18KK0375), Grant-in-Aid for Scientific Research on Innovative Areas ``Quantum Liquid Crystals” (No.\,JP19H05824), Transformative Research Areas (A) ``Condensed Conjugation” (No.\,JP20H05869), and the Graduate School of Frontier Sciences, The University of Tokyo, through the Challenging New Area Doctoral Research Grant (Project No. C2316).
\end{acknowledgments}

\appendix
\section{\label{AP:Derivation}Derivation of the principle for the surface impedance tensor measurements}
In this APPENDIX section, we derive Eq.\,(\ref{resonance_and_Z}) in the main text, which connects the surface impedance tensor to the characteristics of the circularly polarized cavity modes.

\subsection{\label{AP:cavity}Cavity perturbation method}
We overview the conventional cavity perturbation method and derive the ordinary relationship between the resonant properties and the surface impedance of a conductive material.
The discussion in this section is based on Refs.\,\cite{Ogawa_bimodal2021, Slater1946}.

First, we expand the electromagnetic field eigenmodes, $\bm{E}$ and $\bm{H}$, and the current density $\bm{j}$ by the solenoidal vectors, $\bm{e}_a$ and $\bm{h}_a$, and the irrotational vectors, $\bm{f}_a$ and $\bm{g}_a$, where the divergences of $\bm{e}_a$ and $\bm{h}_a$ are zero ($\grad \vdot \bm{e}_a (\bm{h}_a) = 0$) and the rotations of $\bm{f}_a$ and $\bm{g}_a$ are zero ($\grad \times \bm{f}_a (\bm{g}_a) = 0$), as follows:
\begin{equation}
\label{E_expansion}
\bm{E} = \sum_a \left[\bm{e}_a \int_V \dd{\textit{V}} \bm{E} \vdot \bm{e}_a + \bm{f}_a \int_V \dd{V} \bm{E} \vdot \bm{f}_a \right], 
\end{equation}
\begin{equation}
\label{H_expansion}
\bm{H} = \sum_a \left[\bm{h}_a \int_V \dd{\textit{V}} \bm{H} \vdot \bm{h}_a + \bm{g}_a \int_V \dd{V} \bm{H} \vdot \bm{g}_a \right],
\end{equation}
\begin{equation}
\label{j_expansion}
\bm{j} = \sum_a \left[\bm{e}_a \int_V \dd{V} \bm{j} \vdot \bm{e}_a + \bm{f}_a \int_V \dd{V} \bm{j} \vdot \bm{f}_a \right].
\end{equation}
Here $a$ is the index indicating the mode of interest, which is orthonormalized with respect to each of the other modes.
$\bm{e}_a$, $\bm{h}_a$, $\bm{f}_a$, and $\bm{g}_a$ satisfy the following relation and boundary conditions at the surface of a sample and a cavity, 
\begin{eqnarray}
    \label{basis_gene1}
    k_a \bm{e}_a &= \grad \times \bm{h}_a,\,\,\,k_a \bm{h}_a &= \grad \times \bm{e}_a, \\
    \label{basis_gene2}
    k_a \bm{f}_a &= \grad \psi_a,\,\,\,\,\,\,\,\,\,    k_a \bm{g}_a &= \grad \phi_a,
\end{eqnarray}
\begin{eqnarray}
    \bm{n} \times \bm{e}_a = 0, \quad \bm{n} \vdot \bm{h}_a = 0, \quad \psi_a = 0, \quad \phi_a = 0, \label{conduct_boundary}\\
    \bm{n} \times \bm{h}_a = 0, \quad \bm{n} \vdot \bm{e}_a = 0, \quad \psi_a = 0, \quad \phi_a = 0, \label{insulat_boundary}
\end{eqnarray}
where $k_a$ is the propagation constant, and $\psi_a$ and $\phi_a$ are the scalar field for $\bm{f}_a$ and $\bm{g}_a$, respectively.
Eqs.\,(\ref{conduct_boundary}) and (\ref{insulat_boundary}) are the boundary conditions on a conductive and an insulating surface, respectively, and $\bm{n}$ is the unit vector perpendicular to the surface.

Then, we expand Eqs.\,(\ref{E_expansion}), (\ref{H_expansion}), and (\ref{j_expansion}), and substitute them into Maxwell's equations and extract only $a$ component, from which the following equations can be derived
\begin{equation}
    \label{ka_1}
    k_a \int_V \dd{V} \bm{E} \vdot \bm{e}_a + \int_{S_c} \dd{S} (\bm{n} \times \bm{E}) \vdot \bm{h}_a = - \mu_0 \pdv{t} \int_V \dd{V} \bm{H} \vdot \bm{h}_a,
\end{equation}
\begin{equation}
 \begin{split}
    \label{ka_2}
    k_a \int_V \dd{V} \bm{H} \vdot \bm{h}_a + \int_{S_i} \dd{S} (\bm{n} \times \bm{H}) \vdot \bm{e}_a\\
    = \epsilon_0 \pdv{t} \int_V \dd{V} \bm{E} \vdot \bm{e}_a + \int_V \dd{V} \bm{j} \vdot \bm{e}_a,
 \end{split}
\end{equation}
where $S_c$ is the conducting surface, $S_i$ is the insulating surface, and $V$ is the volume of the cavity. 
We arranged these equations as follows:
\begin{equation}
  \begin{split}
   \label{em_cavity_Ebasis_eq}
    \bigl(k_a^2 + \epsilon_0 \mu_0 & \pdv[2]{t} \bigr) \int_V \dd{V} \bm{E} \vdot \bm{e}_a \\
    &= -k_a \int_{S_c} \dd{S} (\bm{n} \times \bm{E}) \vdot \bm{h}_a\\
    &+ \mu_0 \pdv{t} \int_{S_i} \dd{S} (\bm{n} \times \bm{H}) \vdot \bm{e}_a - \mu_0 \pdv{t} \int_V \dd{V} \bm{j} \vdot \bm{e}_a,
  \end{split}
\end{equation}
\begin{equation}
  \begin{split}
    \label{em_cavity_Hbasis_eq}
    \bigl(k_a^2 + \epsilon_0 \mu_0 & \pdv[2]{t} \bigr) \int_V \dd{V} \bm{H} \vdot \bm{h}_a \\
    = &- \epsilon_0 \pdv{t} \int_{S_c} \dd{S} (\bm{n} \times \bm{E}) \vdot \bm{h}_a\\
     &- k_a \int_{S_i} \dd{S} (\bm{n} \times \bm{H}) \vdot \bm{e}_a + k_a \int_V \dd{V} \bm{j} \vdot \bm{e}_a.\,\,\,\,\,\,
  \end{split}
\end{equation}
Here, in an ideal cavity surrounded by a perfect conducting surface, the right-hand side of Eq.\,(\ref{em_cavity_Ebasis_eq}) is zero.
Then, we can obtain the following relation for the dispersion of the electromagnetic field by considering the time revolution of $e^{i\omega t}$,
\begin{equation}
    \label{dispersion_relation}
    \omega_a = \dfrac{1}{\sqrt{\epsilon_0 \mu_0}} k_a .
\end{equation}
From this equation, Eq.\,(\ref{em_cavity_Hbasis_eq}) is rewritten by
\begin{equation}
  \begin{split}
    \label{em_cavity_Hbasis_eq_2}
    i(\frac{\omega}{\omega_a}-\frac{\omega_a}{\omega}) \int_V \dd{V} \bm{H} \vdot \bm{h}_a 
    = -\frac{1}{\omega_a \mu_0} \int_{S_c} \dd{S} (\bm{n} \times \bm{E}) \vdot \bm{h}_a\\
     + \frac{i}{\omega\sqrt{\epsilon_0 \mu_0}} (\int_{S_i} \dd{S} (\bm{n} \times \bm{H}) \vdot \bm{e}_a -  \int_V \dd{V} \bm{j} \vdot \bm{e}_a).
  \end{split}
\end{equation}
An actual cavity has dissipation, and the complex angular resonance frequency $\hat \omega$ of the cavity in the mode $a$ can be expressed using the deviation $\delta \hat \omega_a$ from the ideal cavity and the $Q$-factor as follows:
\begin{equation}
    \hat \omega = \hat \omega_a + \delta \hat \omega_a \simeq \omega_a + \dfrac{i \omega_a}{2Q} + \delta \omega_a. \label{omega_cav}
\end{equation}
If the dissipation is perturbative ($Q\gg1$), we can obtain from Eq.\,(\ref{omega_cav})
\begin{equation}
    \label{shift_Q_omega}
    \dfrac{i}{2} \Bigl( \dfrac{\omega}{\omega_a} - \dfrac{\omega_a}{\omega} \Bigr) \simeq -\dfrac{1}{2Q} + i\dfrac{\delta \omega_a}{\omega_a}.
\end{equation}
From Eqs.\,(\ref{em_cavity_Hbasis_eq_2}) and (\ref{shift_Q_omega}), we can obtain the following relationship:
\begin{align}
    \label{em_A}
    \begin{split}
    -\dfrac{1}{2Q} &+ i\dfrac{\delta \omega_a}{\omega_a}
    \simeq -\dfrac{\int_{S_c} \dd{S} (\bm{n} \times \bm{E}) \vdot \bm{h}_a}{2 \mu_0\omega_a \int_V \dd{V} \bm{H} \vdot \bm{h}_a}\\ 
    &+ \dfrac{i}{2 \omega Z_0 \epsilon_0} \Bigl(\dfrac{\int_{S_i} \dd{S} (\bm{n} \times \bm{H}) \vdot \bm{e}_a}{\int_V \dd{V} \bm{H} \vdot \bm{h}_a} - \dfrac{\int_V \dd{V} \bm{j} \vdot \bm{e}_a}{\int_V \dd{V} \bm{H} \vdot \bm{h}_a} \Bigr ),
    \end{split}
\end{align}
where $Z_0 = \sqrt{\mu_0 / \epsilon_0}$ represents the vacuum impedance.
Here, we consider the perturbation response when a small conducting sample is placed in the antinode of the magnetic field mode in the cavity.
The second term on the right-hand side of Eq.\,(\ref{em_A}) indicates the Joule heating contribution due to the resistance of the sample and the cavity walls.
If the sample is sufficiently smaller than the cavity volume, the contribution from the sample is negligible. 
The third term is the contribution from the insulating surface.
Therefore, only the first term contributes to the response of the conducting sample.
Then, considering the difference $\Delta$ with and without the sample, the second and third terms can be subtracted as background contributions, and we can obtain the following essential relationship between the magnetic field mode and the frequency properties in the cavity perturbation method as follows:
\begin{align}
    \label{cav_p_base}
    \begin{split}
    \Delta \Bigl( \dfrac{1}{2Q} - i\dfrac{\delta \omega_a}{\omega_a} \Bigr) &= \Delta \Bigl( \dfrac{1}{2Q} \Bigr) - i \Bigl( \dfrac{\Delta \omega}{\omega_a} + C \Bigr)\\ 
    &=  \dfrac{\int_S \dd{S} (\bm{n} \times \bm{E}) \vdot \bm{h}_a}{2 \mu_0 \omega_a \int_V \dd{V} \bm{H} \vdot \bm{h}_a}.
    \end{split}
\end{align}
\subsection{\label{AP:impedance}Principle of the surface impedance tensor measurements}
In this section, we extend Eq.\,(\ref{cav_p_base}) to the case of circularly polarized magnetic field modes.
First, we introduce the degree of freedom for the right\,(+) and left\,(-)-handed polarized modes into Eq.\,(\ref{cav_p_base}) as follows:
\begin{equation}
   \label{cavity_basic_formula_CP}
    \Delta \Bigl(\frac{1}{2 Q_{\pm}}\Bigr) - i\Bigl(\frac{\Delta \omega_{\pm}}{\omega_0} + C\Bigr) = {\dfrac{\int_S \dd{S} (\bm{n} \times \bm{E}) \vdot \bm{h}_a}{2 \mu_0 \omega_a \int_V \dd{V} \bm{H} \vdot \bm{h}_a}},
\end{equation}
where $\bm{H_a}$ is the magnetic field of the circularly polarized mode we are currently focusing on.
Next, we get a specific expression for $ \int_{S} \dd{S} (\bm{n} \cross \bm{E}) \vdot \bm{H_a}$.
Let us consider the basis expansion of the right\,(+) and left\,(-)-handed circularly polarized modes in the $xy$-plane as shown in Sec.\,\ref{sec:2-A}.
They are described as
\begin{equation}
  \label{CP_base}
   \bm{H}_a = h\mqty(1\\0) + h\mqty(0\\ \pm i),
\end{equation}
where $h$ is the amplitude of the magnetic field in the mode.
We decompose the circularly polarized mode $H_a$ into two linearly polarized basic vectors with a phase difference of $\pm\frac{\pi}{2}$ as follows:
\begin{subequations}
\begin{eqnarray}
\bm{H}_{\rm{A}}
 &= &h \mqty(1\\0), \label{basic_v_linear_A}
\\
\bm{H}_{\rm{B}} &= &h \mqty(0\\ \pm i). \label{basic_v_linear_B}
\end{eqnarray}
\end{subequations}
Then, $\bm{H}_a$ can be expressed as
\begin{equation}
\begin{split}
\bm{H}_a &\thickapprox \bm{H}_{\rm{A}} \int_V \bm{H} \vdot \bm{H}_{\rm{A}} \dd{V} + \bm{H}_{\rm{B}} \int_V \bm{H} \vdot \bm{H}_{\rm{B}} \dd{V} \\
&= h \mqty(\int_V \bm{H} \vdot \bm{H}_{\rm{A}} \dd{V}\\ \pm i \int_V \bm{H} \vdot \bm{H}_{\rm{B}} \dd{V}).
\end{split}
\label{mode_expansion}
\end{equation}
Next, we consider the specific form of $(\bm{n} \cross \bm{E}) \vdot \bm{H}_a$.
In this case, the ratio of the electric and magnetic field components parallel to the sample surface is defined by the surface impedance tensor from Eq.\,(\ref{Z_tensor}) as, 
\begin{equation}
  \begin{split}
    \bm{n} \cross \bm{E} &= \tilde Z \bm{H} \\
     &= \tilde Z \bm{H}_a \\
     &= h \mqty(Z_{\rm{L}} \int_V \bm{H} \vdot \bm{H}_{\rm{A}} \dd{V} \pm i Z_{\rm{H}} \int_V \bm{H} \vdot \bm{H}_{\rm{B}} \dd{V} \\ -Z_{\rm{H}} \int_V \bm{H} \vdot \bm{H}_{\rm{A}} \dd{V} \pm i Z_{\rm{L}} \int_V \bm{H} \vdot \bm{H}_{\rm{B}} \dd{V}).
    \end{split}
    \label{nxE_fomula}
\end{equation}
Here, we consider that the component of the total magnetic field $H$ parallel to the sample surface ($xy$-plane), which contributes to the surface impedance, is $\bm{H}_a$.
From Eqs.\,(\ref{CP_base}) and (\ref{nxE_fomula}), $(\bm{n} \cross \bm{E}) \vdot \bm{H}_a$ is expressed as
\begin{align}
\label{nxE_dot_H_a_formula}
 \begin{split}
  \left(\bm{n} \cross \bm{E} \right) \vdot \bm{H}_a &=
   h^2  \biggl\{Z_{\rm{L}} \left(\int_V \bm{H} \vdot \bm{H}_{\rm{A}} \dd{V} - \int_V \bm{H} \vdot \bm{H}_{\rm{B}} \dd{V} \right)\\
   &\mp i Z_{\rm{H}} \left(\int_V \bm{H} \vdot \bm{H}_{\rm{A}} \dd{V} - \int_V \bm{H} \vdot \bm{H}_{\rm{B}} \dd{V}\right) \biggr\}.
 \end{split}
\end{align}
The substitution of Eq.\,(\ref{nxE_dot_H_a_formula}) into Eq.\,(\ref{cavity_basic_formula_CP}) leads to the following relationship between the resonant properties (frequency and $Q$-factor) and the surface impedance tensor
\begin{gather}
  \label{f_Q_shift_surface_impedance_formula}
  \begin{split}
   \Delta\Bigl(\frac{1}{2 Q_{\pm}}\Bigr) - i\Bigl(\frac{\Delta \omega_{\pm}}{\omega_0} + C\Bigr)
   = \frac{\int_{S} h^2 \dd{S}}{2\omega_0 \mu_0 \int_V \bm{H} \vdot \bm{H}_{\rm{A}} \dd{V}}\\
   \cross \biggl\{ Z_{\rm{L}} \Bigl(\int_V \bm{H} \vdot \bm{H}_{\rm{A}} \dd{V} - \int_V \bm{H} \vdot \bm{H}_{\rm{B}} \dd{V} \Bigr) \\
   \mp i Z_{\rm{H}} \Bigl(\int_V \bm{H} \vdot \bm{H}_{\rm{A}} \dd{V} - \int_V \bm{H} \vdot \bm{H}_{\rm{B}} \dd{V} \Bigr) \biggr\} \\
   = G \left(Z_{\rm{L}} \mp i Z_{\rm{H}}\right),\,\,\,\,\,\,\,\,\,\,\,\,\,\,\,\,\,\,\,\,\,\,\,\,\,\,\,
   \end{split}
\end{gather}
where $\omega_0$ is the resonance frequency without the sample, and the geometric factor $G$ is defined by
\begin{equation}
\label{geometric_factor}
   G = \frac{\int_{S} h^2 ds}{2\omega_0 \mu_0 \int_V \bm{H} \vdot \bm{H}_{\rm{A}} \dd{V}} \Bigl(\int_V \bm{H} \vdot \bm{H}_{\rm{A}} \dd{V} - \int_V \bm{H} \vdot \bm{H}_{\rm{B}} \dd{V} \Bigr).
\end{equation}
As a result, we derive Eq.\,(\ref{resonance_and_Z}) in the main text, which connects the surface impedance tensor to the characteristics of the circularly polarized cavity modes.

\begin{figure}[t]
\hspace*{-1.0cm}
\includegraphics[width=0.9\linewidth, clip]{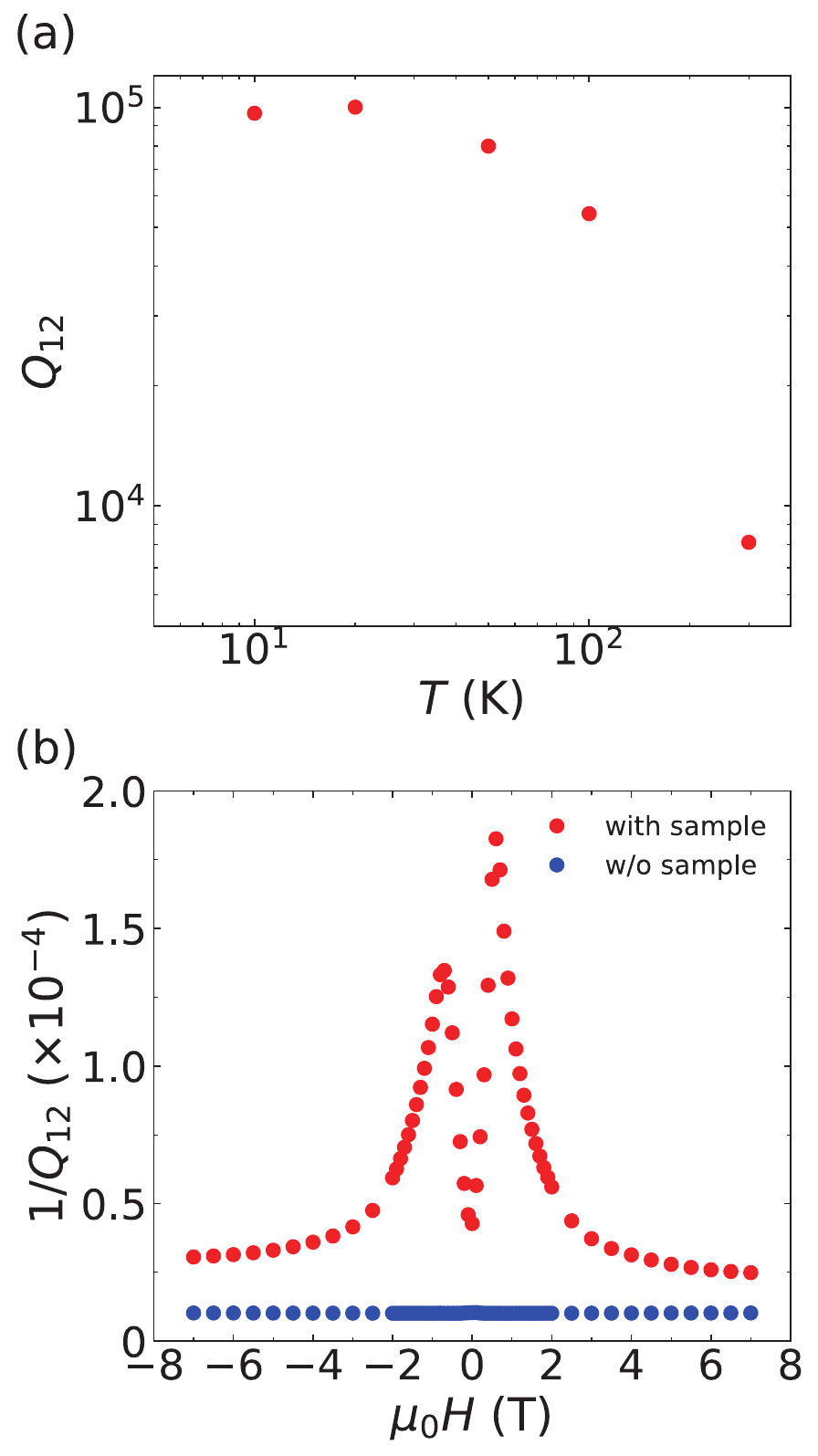}
\caption{Background data for the $Q$-value of the circularly polarized rutile cavity used in this study. (a) Temperature dependence of $Q_{12}$ without a sample in a zero magnetic field. (b) Magnetic field dependence of $1/Q_{12}$ with (red) and without (blue) a sample measured at 10\,K.}
\label{fig:Q(T)}
\end{figure}

\section{\label{AP_DC}Details of dc transport measurements}
Our dc transport measurements were performed using a five-terminal method with PPMS, applying magnetic fields in both the positive and negative directions along the trigonal axis of Bi.
We obtained $\rho_{xx}$ and $\rho_{yx}$, and calculated $\sigma_{xx}$ and $\sigma_{xy}$ by using the following relations:
\begin{eqnarray}
    \sigma_{xx} = \frac{\rho_{xx}}{{\rho_{xx}}^2 + {\rho_{yx}}^2}, \label{sigma_xx_invert}\\
    \sigma_{xy} = \frac{\rho_{yx}}{{\rho_{xx}}^2 + {\rho_{yx}}^2}. \label{sigma_xy_invert}
\end{eqnarray}
To investigate the validity of our measurements and the quality of our Bi sample, we fitted $\sigma_{xx}$ and $\sigma_{xy}$ with the following equations in the two-carrier model.
\begin{eqnarray}
  \sigma_{xx} &=& \dfrac{n_{\rm{e}} e \mu_{\rm{e}}}{1 + (\mu_{\rm{e}} B)^2} + \dfrac{n_{\rm{h}} e \mu_{\rm{h}}}{1 + (\mu_{\rm{h}} B)^2},   \label{two_career_xx} \\
  \sigma_{xy} &=& -\dfrac{n_{\rm{e}} e \mu_{\rm{e}}^2 B}{1 + (\mu_{\rm{e}} B)^2} + \dfrac{n_{\rm{h}} e \mu_{\rm{h}}^2 B}{1 + (\mu_{\rm{h}} B)^2},\label{two_career_xy}
\end{eqnarray}
where $n$ is the carrier density, $\mu$ is the mobility, $e$ is the elementary charge, $B$ is the magnetic field, and the subscripts $\rm{e}$ and $\rm{h}$ are indexes for electron and hole carriers, respectively.

\section{\label{Q_bg} $Q$-factor of circularly polarized rutile cavity}
We measured the temperature and magnetic field dependence of $Q_{12}$ for the circularly polarized rutile cavity used in this study from background measurements without a sample. Figure\,\ref{fig:Q(T)}(a) presents the temperature dependence of $Q_{12}$ without a sample in a zero magnetic field. At room temperature, the $Q$-factor is low ($\sim 8\times10^3$) due to the small dielectric constant of rutile, but it increases as the temperature decreases and reaches $\sim 1\times10^5$ at low temperatures. Figure\,\ref{fig:Q(T)}(b) shows the magnetic field dependence of $1/Q_{12}$ with and without a sample measured at 10\,K. The background variation of $1/Q_{12}$ with respect to the magnetic field is considerably smaller than the variation of the sample.

\section*{AUTHOR DECLARATIONS}
\subsection*{Data Availability Statement}
The data that support the findings of this study are available from the corresponding authors upon reasonable request.

\subsection*{Conflict of Interest}
The authors have no conflicts to disclose.

\section*{REFERENCES}
\nocite{*}
\bibliography{circular_cavity}

\end{document}